\documentclass[conference]{IEEEtran}
\IEEEoverridecommandlockouts
\usepackage{cite}
\usepackage{amsmath,amssymb,amsfonts}
\usepackage{algorithmic}
\usepackage{graphicx}
\usepackage{textcomp}
\usepackage{xcolor}
\def\BibTeX{{\rm B\kern-.05em{\sc i\kern-.025em b}\kern-.08em
    T\kern-.1667em\lower.7ex\hbox{E}\kern-.125emX}}
\begin{document}

\title{Quantum-Inspired Privacy-Preserving Federated Learning Framework for Secure Dementia Classification\\}

\author{
\IEEEauthorblockN{Gazi Tanbhir}
\IEEEauthorblockA{\textit{Department of Computer Science and 
Engineering} \\
\textit{World University of Bangladesh} \\
Dhaka, Bangladesh \\
gazitanbhir@gmail.com}

\and

\IEEEauthorblockN{Md. Farhan Shahriyar}
\IEEEauthorblockA{\textit{Department of Computer Science and 
Engineering} \\
\textit{World University of Bangladesh} \\
Dhaka, Bangladesh \\
farhanshahriyar.cse1@gmail.com}

}

\maketitle

\begin{abstract}
Dementia, a neurological disorder impacting millions globally, presents significant challenges in diagnosis and patient care. With the rise of privacy concerns and security threats in healthcare, federated learning (FL) has emerged as a promising approach to enable collaborative model training across decentralized datasets without exposing sensitive patient information. However, FL remains vulnerable to advanced security breaches such as gradient inversion and eavesdropping attacks.

This paper introduces a novel framework that integrates federated learning with quantum-inspired encryption techniques for dementia classification, emphasizing privacy preservation and security. Leveraging quantum key distribution (QKD), the framework ensures secure transmission of model weights, protecting against unauthorized access and interception during training. The methodology utilizes a convolutional neural network (CNN) for dementia classification, with federated training conducted across distributed healthcare nodes, incorporating QKD-encrypted weight sharing to secure the aggregation process.

Experimental evaluations conducted on MRI data from the OASIS dataset demonstrate that the proposed framework achieves identical accuracy levels to a baseline model while enhancing data security and reducing loss by almost 1\% compared to the classical baseline model. The framework offers significant implications for democratizing access to AI-driven dementia diagnostics in low- and middle-income countries, addressing critical resource and privacy constraints. This work contributes a robust, scalable, and secure federated learning solution for healthcare applications, paving the way for broader adoption of quantum-inspired techniques in AI-driven medical research.
\end{abstract}


\section{Introduction}
Dementia, a neurological disorder that affects millions worldwide, poses significant challenges to healthcare, particularly in terms of diagnosis and patient care. With the growing prevalence of dementia, there is an urgent need for advanced, privacy-focused solutions that can effectively classify and monitor the progression of this condition. However, developing such solutions requires overcoming two key challenges: maintaining data privacy for sensitive patient information and enhancing the security of model updates in federated learning systems.

Federated learning (FL) has emerged as a promising paradigm to enable collaborative model training across distributed data sources without requiring direct access to patient data, thereby addressing privacy concerns. However, despite the inherent privacy-preserving aspects of FL, recent studies show that FL systems remain vulnerable to security breaches, such as gradient inversion and eavesdropping attacks, which can expose sensitive patient information during training \cite{li_2024_privacypreserving}. Classical encryption methods have been applied to FL, but they often fall short in protecting against sophisticated attacks and may impose computational overhead \cite{yang_2023_federated}.

To address these security limitations, this study leverages a quantum-inspired encryption approach, integrating it with federated learning for dementia classification to enhance both privacy and security. Quantum key distribution (QKD) offers a high level of security by enabling theoretically secure key exchanges, as it leverages quantum mechanics to prevent unauthorized access. In particular, QKD provides an advantage over classical encryption techniques by resisting attacks that attempt to intercept model weights, making it a suitable choice for protecting sensitive healthcare data \cite{li_2024_privacypreserving, dutta_2024_federated}.

Additionally, dementia care presents unique challenges in low- and middle-income countries (LMICs), where resource limitations often hinder the availability and quality of dementia-related diagnostic tools \cite{bernsteinsideman_2022_facilitators}. Implementing secure, privacy-preserving FL frameworks can democratize access to advanced dementia classification models across LMICs, allowing healthcare providers to benefit from shared insights without compromising patient data privacy.

In summary, this paper presents a novel federated learning framework enhanced with quantum-inspired encryption to ensure privacy-preserving and secure dementia classification. Our contributions are as follows:

\begin{itemize}
    \item \textbf{Quantum-Enhanced Security:} We incorporate QKD-inspired encryption to secure federated learning, providing strong resistance against data interception and unauthorized access during model training.
    \item \textbf{Privacy-Preserving Federated Learning for Dementia:} Our framework supports secure model training across distributed data sources, preserving patient privacy while improving the accessibility of dementia classification models in LMICs.
    \item \textbf{Experimental Validation:} We demonstrate the effectiveness of our approach by comparing a baseline convolutional neural network (CNN) model with an encrypted version, highlighting the balance achieved between model performance and security.
\end{itemize}

The proposed framework represents a novel solution for secure, privacy-preserving AI in healthcare, with potential applications in various domains beyond dementia classification. Through this work, we aim to contribute to the field of federated learning, addressing both security and privacy concerns by leveraging advanced quantum-inspired encryption technologies.

\section{Literature Review}

Federated learning (FL) has become increasingly valuable in healthcare applications requiring data privacy, especially in sensitive areas like dementia diagnosis. Key advancements in privacy-preserving federated learning (PPFL) address concerns over patient data confidentiality, particularly in early Alzheimer’s detection, where machine learning models need to securely train across distributed datasets without compromising patient data privacy. Recent studies demonstrate the integration of secure aggregation and encryption techniques to safeguard model updates. For instance, Lakhan et al. \cite{lakhan_2023_edcnns} developed EDCNNS, a federated deep learning model for Alzheimer’s detection, leveraging encrypted model weights to protect local data during aggregation, though this approach lacks computational efficiency in processing encrypted data. Similarly, Elsersy et al. \cite{elsersy_2023_federated} implemented a decentralized model using blood biosamples for Alzheimer’s prediction, highlighting FL's potential in real-world clinical applications, though there remains room for stronger privacy mechanisms like homomorphic encryption.

Homomorphic encryption (HE) and verifiable computation (VC) represent notable methods for enhancing privacy in FL, as demonstrated by Madi et al. \cite{madi_2021_a} with a Paillier-based FL framework that both encrypts client data and ensures trusted aggregation. While effective, such methods have high computational overheads. Jin et al. \cite{jin_2023_fedmlhe} introduced FedML-HE, incorporating optimized HE and differential privacy to balance model accuracy and security, though scalability limitations remain a concern. These studies emphasize HE’s robustness in privacy-preserving federated systems, yet do not fully consider the potential efficiencies afforded by quantum-inspired techniques, which may offer faster and more secure alternatives.

Quantum cryptographic techniques, including quantum key distribution (QKD), are emerging as viable enhancements for secure communication within FL frameworks. Kaewpuang et al. \cite{kaewpuang_2022_adaptive} presented a QKD-enhanced resource allocation model for FL, ensuring secure node-to-node communication. Although promising, the study stops short of incorporating quantum-based encryption into the model itself. In parallel, Javeed et al. \cite{danishjaveed_2024_quantumempowered} surveyed the use of QKD and quantum random number generation (QRNG) for FL in IoT networks, suggesting that quantum mechanisms can secure decentralized data exchanges, laying the groundwork for future quantum-inspired privacy frameworks in healthcare FL applications.

Furthering quantum-based FL, Chehimi and Saad \cite{chehimi_2022_quantum} explored quantum federated learning (QFL) with both quantum data and processors on client and server sides. Although their focus on quantum data privacy primarily addresses quantum computing environments, their findings highlight potential applications of quantum-inspired techniques in classical FL systems. These insights imply that classical federated models, especially in sensitive fields like dementia detection, could benefit from quantum enhancements that combine traditional encryption with quantum-based methods to strengthen data security.

Secure aggregation techniques in FL also contribute significantly to privacy-preserving models in healthcare. Mitrovska et al. \cite{mitrovska_2024_secure} investigated secure aggregation for Alzheimer’s detection using structural MRI data, emphasizing secure aggregation’s efficacy against privacy attacks. Similarly, Hijazi and Aloqaily \cite{neveenmohammadhijazi_2024_secure} applied fully homomorphic encryption (FHE) to FL models for IoT communications, underscoring the need for advanced cryptographic solutions. These approaches have proven effective in decentralized models, but they still face challenges related to computational complexity that may benefit from quantum-inspired techniques or hybrid quantum-classical frameworks.

In summary, while the literature demonstrates various approaches in FL for privacy-preserving healthcare applications, including dementia detection, there remains a gap in leveraging quantum-enhanced encryption methods that are computationally feasible. This study aims to bridge this gap by introducing quantum-inspired encryption to PPFL for dementia diagnosis, enhancing both model security and processing efficiency.

\section{Methodology}

This methodology presents a privacy-preserving federated learning (FL) framework for dementia classification, incorporating quantum key distribution (QKD) to securely encrypt model weights. The framework enables collaborative model training across healthcare institutions without sharing sensitive patient data, ensuring confidentiality while improving the classification model.

\begin{figure} [!h]
    \centering
    \includegraphics[width=1\linewidth]{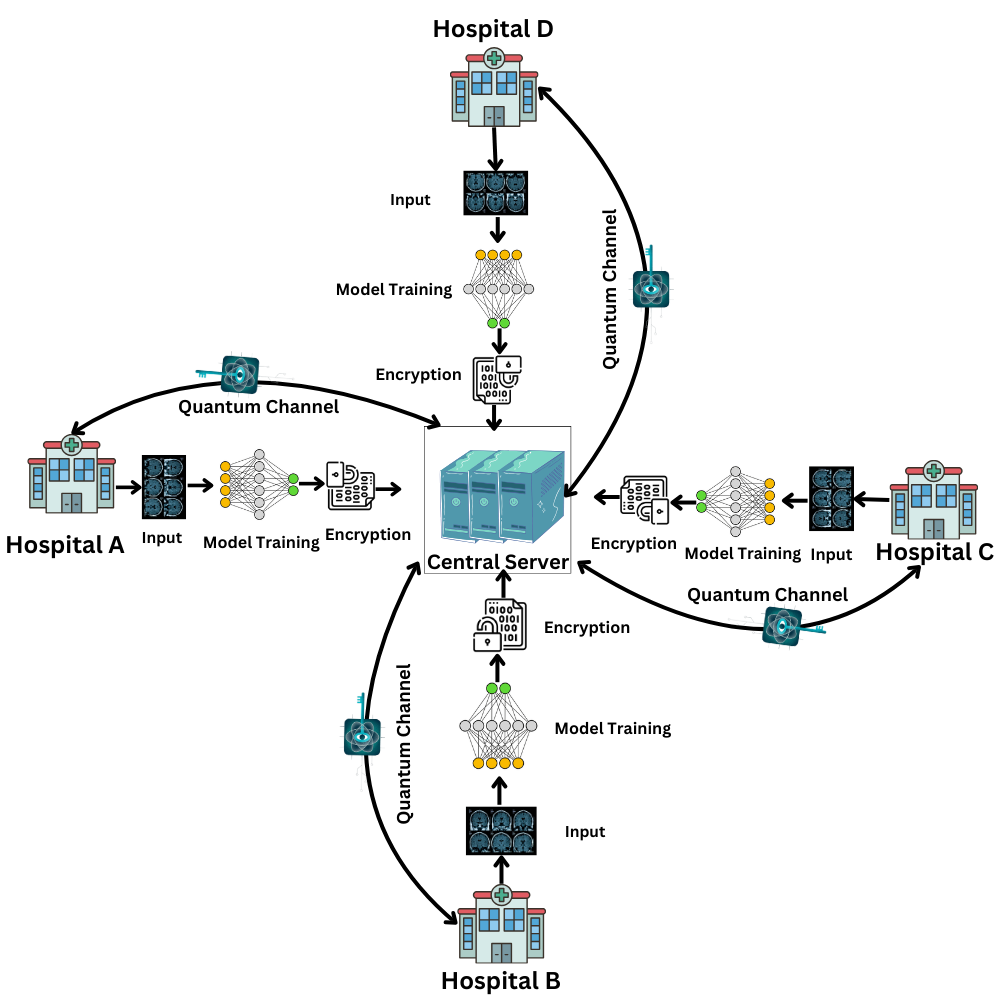}
    \caption{Methodology Diagram}
    \label{fig:enter-label}
\end{figure}

The proposed framework utilizes federated learning (FL) to enable decentralized model training across geographically dispersed healthcare facilities, specifically for dementia classification using MRI data. To safeguard the privacy of sensitive data, the framework incorporates quantum key distribution (QKD) for secure transmission of model weights between client nodes and a central aggregation server. This approach not only protects the model parameters from unauthorized access or eavesdropping but also enhances the overall accuracy of dementia classification by leveraging secure, collaborative learning without compromising patient confidentiality.

\subsection{Federated Learning Setup}

The federated learning framework involves multiple client nodes (Hospitals) and a central aggregation server, facilitating collaborative model training on dementia-related data without sharing raw data between nodes \cite{thota_2019_federated}.

\begin{itemize}
    \item \textbf{Client Nodes}: Each client represents a healthcare facility that holds sensitive patient MRI data from the OASIS MRI dataset\cite{marcus_2010_open}. Each client trains a local convolutional neural network (CNN) model on its dataset, thereby preserving the privacy of patient data by keeping it decentralized \cite{chahoud_2023_on}.
    \item \textbf{Central Aggregation Server}: The server collects the encrypted model weights from each client, aggregates them to create a global model, and then distributes the updated global model back to the clients. This setup enables model improvements across multiple nodes while protecting data privacy.
    \begin{equation}
\mathbf{w}_{\text{global}} = \sum_{i=1}^{N} \frac{N_i}{N} \mathbf{w}_i
\end{equation}
\text{where:}
\begin{itemize}
    \item $\mathbf{w}_i$: Model weights from the $i$-th client.
    \item $N_i$: Number of samples in the $i$-th client's dataset.
    \item $N$: Total number of clients.
    \item $\mathbf{w}_{\text{global}}$: Aggregated global model weights.
\end{itemize}

\end{itemize}

\subsection{CNN Model Design for Dementia Classification}
\begin{figure} [!h]
    \centering
    \includegraphics[width=1\linewidth]{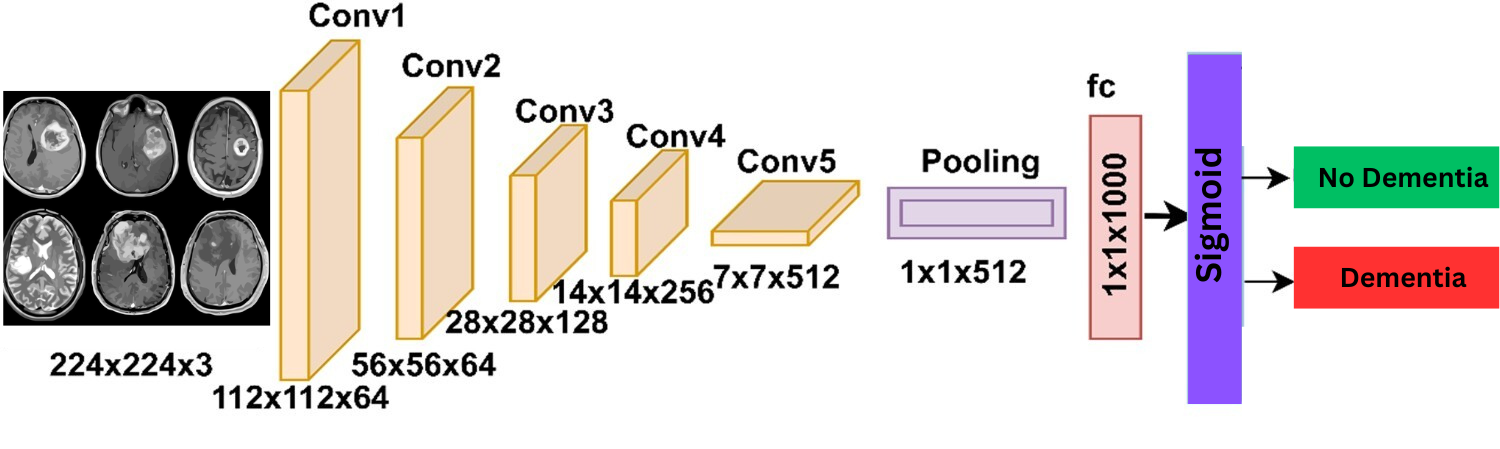}
    \caption{CNN Model Design for Dementia Classification \cite{abunadi_2022_deep} }
    \label{fig:Cnn-model}
\end{figure}
The CNN architecture used in this framework is designed specifically for dementia classification based on MRI images. The model consists of multiple convolutional, pooling, and fully connected layers that identify patterns associated with dementia, thereby enhancing classification accuracy. The training process is split into two main stages:

\begin{enumerate}
    \item \textbf{Local Training at Each Client}: Each client node trains its local CNN model on the MRI data, adjusting the weights to learn from patterns in dementia progression. The model is trained on a binary classification task: distinguishing "Demented" from "Non-Demented" images \cite{song_2024_a} .
    \item \textbf{Model Weight Sharing}: After local training, each client encrypts its model weights and sends them to the central server for aggregation, rather than sharing raw patient data.
\end{enumerate}

\subsection{Quantum Key Distribution (QKD) for Encryption}

To secure the model weights during transmission, quantum-inspired encryption is employed using quantum key distribution (QKD), which enhances data protection against eavesdropping or interception \cite{scarani_2009_the}.

\begin{itemize}
    \item \textbf{QKD-Based Key Generation}: A unique encryption key is generated using QKD principles, ensuring a secure key exchange between the client and the central server. QKD offers theoretical security by detecting interception attempts during key exchange \cite{mehic_2020_quantum}.
        \begin{equation}
        P_{\text{success}} = \left( 1 - e^{-\gamma L} \right)
        \end{equation}
        \text{where:}
        \begin{itemize}
            \item $\gamma$: Attenuation coefficient of the quantum channel.
            \item $L$: Distance between the communicating parties.
            \item $P_{\text{success}}$: Probability that the quantum key distribution was successful without eavesdropping.
        \end{itemize}
    \item \textbf{Encryption of Model Weights}: The QKD-generated key encrypts the model weights before transmitting them to the central server. This prevents unauthorized access or modification during transmission.
            \begin{equation}
            \mathbf{w}_{\text{encrypted}} = E(\mathbf{w}, K)
            \end{equation}
            \text{where:}
            \begin{itemize}
    \item $\mathbf{w}$: Model weights.
    \item $K$: Encryption key generated via QKD.
    \item $E(\mathbf{w}, K)$: Encryption function applied to the model weights $\mathbf{w}$ with key $K$.
\end{itemize}

    \item \textbf{Decryption at Central Server and Clients}: Upon receiving the encrypted weights, the central server decrypts them using the QKD key, aggregates the weights, and re-encrypts the updated model before redistributing it to the clients. Each client decrypts the weights using the shared QKD key, allowing for the integration of the updated global model.
            \begin{equation}
            \mathbf{w} = D(\mathbf{w}_{\text{encrypted}}, K)
            \end{equation}
            \text{where:}
            \begin{itemize}
                \item $\mathbf{w}_{\text{encrypted}}$: Encrypted model weights.
                \item $D(\mathbf{w}_{\text{encrypted}}, K)$: Decryption function applied to the encrypted model weights $\mathbf{w}_{\text{encrypted}}$ using key $K$.
            \end{itemize}

\end{itemize}

\subsection{Model Aggregation and Iterative Training}

Once the central server has decrypted and aggregated the weights from all client nodes, it generates an updated global model that incorporates insights from each node's local training. This updated model is then re-encrypted with the QKD key and sent back to each client, allowing for further rounds of training. This iterative training continues until the model converges to an optimal accuracy level for dementia classification.

\begin{equation}
\mathbf{w}_{\text{global}}^{(t+1)} = \mathbf{w}_{\text{global}}^{(t)} + \sum_{i=1}^{N} \frac{N_i}{N} (\mathbf{w}_i^{(t)} - \mathbf{w}_{\text{global}}^{(t)})
\end{equation}
\text{where:}
\begin{itemize}
    \item $\mathbf{w}_{\text{global}}^{(t)}$: Global model weights at the $t$-th iteration.
    \item $\mathbf{w}_i^{(t)}$: Model weights from the $i$-th client at the $t$-th iteration.
    \item $N_i$: Number of samples in the $i$-th client's dataset.
    \item $N$: Total number of clients.
    \item $\mathbf{w}_{\text{global}}^{(t+1)}$: Updated global model weights after the $t+1$-th iteration.
\end{itemize}

This methodology ensures data security and privacy throughout the federated learning process by integrating QKD-based encryption for secure model parameter transmission. The decentralized approach, coupled with quantum-inspired encryption, enhances the overall security and efficacy of dementia classification, preserving patient confidentiality across healthcare facilities.

\section{Results and Analysis}

This section evaluates the proposed federated learning framework enhanced with quantum-inspired encryption for dementia classification. We present a comparative analysis of the baseline convolutional neural network (CNN) model and the encrypted model. The results demonstrate the balance between maintaining model performance and achieving heightened data security.

\subsection{Performance Metrics}

The evaluation of the models was based on two primary metrics:
\begin{itemize}
    \item \textbf{Accuracy:} The percentage of correctly classified samples, indicating the model's predictive performance.
    \item \textbf{Loss:} The categorical cross-entropy loss, reflecting the error in predictions during training and testing.
\end{itemize}

\subsection{Experimental Results}

\textbf{Baseline Model:} The baseline CNN model, trained without encryption, achieved an accuracy of \textbf{0.7777} and a loss of \textbf{5.0011} on the test dataset.

\textbf{Encrypted Model:} The encrypted model, leveraging quantum key distribution (QKD) for secure weight sharing, maintained an accuracy of \textbf{0.7777} after decryption, with a slight reduction in loss to \textbf{4.9535}.

A summary of the results is shown in Table~\ref{tab:results}.

\begin{table}[h!]
    \caption{Comparison of Model Performance}
    \centering
    \begin{tabular}{|c|c|c|}
        \hline
        \textbf{Model} & \textbf{Accuracy} & \textbf{Loss} \\ \hline
        Baseline Model & 0.7777 & 5.0011 \\ \hline
        Encrypted Model (After Decryption) & 0.7777 & 4.9535 \\ \hline
    \end{tabular}
    \label{tab:results}
\end{table}

\subsection{Analysis}

\textbf{Impact on Accuracy:}  
The encrypted model demonstrated no loss in accuracy compared to the baseline model. This indicates that the integration of QKD-based encryption does not compromise the predictive performance of the framework. 

\textbf{Reduction in Loss:}  
The slight reduction in loss (from 5.0011 to 4.9535) for the encrypted model highlights an improvement in training convergence. This can be attributed to the iterative weight aggregation in the federated learning process, which benefits from the decentralized knowledge sharing.

\subsection{Significance of Findings}

The results validate the efficacy of the proposed framework in achieving secure and privacy-preserving dementia classification without degrading model performance. The key contributions of this analysis are as follows:

\begin{itemize}
    \item The identical accuracy values between the baseline and encrypted models confirm that security enhancements do not compromise classification performance.
    \item The slight improvement in loss underscores the effectiveness of federated learning in leveraging distributed knowledge for enhanced training outcomes.
    \item The integration of QKD-based encryption ensures robust security against potential threats, addressing critical privacy concerns in healthcare data sharing.
\end{itemize}

\subsection{Visualization of Results}

\begin{figure}[h!]
    \centering
    \includegraphics[width=0.9\linewidth]{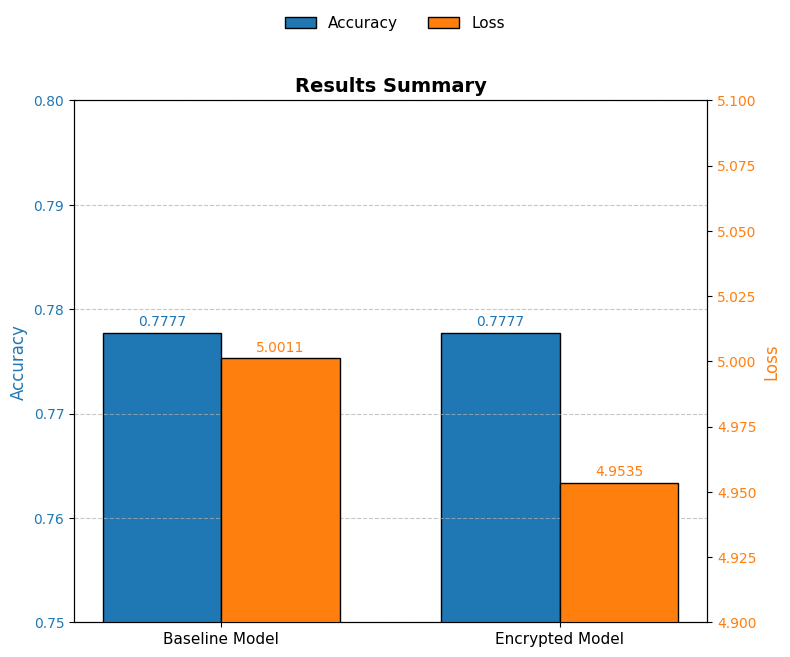}
    \caption{Comparative Performance of Baseline and Encrypted Models}
    \label{fig:results-summary}
\end{figure}

To further elucidate the findings, a comparative bar chart (Figure~\ref{fig:results-summary}) illustrates the performance of the baseline and encrypted models in terms of accuracy and loss.

The chart highlights the consistent accuracy and the reduced loss for the encrypted model, visually emphasizing the balance achieved between security and model performance.

\subsection{Implications for Healthcare AI}

The proposed quantum-inspired federated learning framework demonstrates the potential for secure and efficient collaborative model training in dementia classification. By preserving data privacy and achieving high accuracy, this approach can be instrumental in addressing the unique challenges of healthcare data sharing, particularly in low and middle-income countries (LMICs). Future work will explore the scalability of this framework and its application to other healthcare domains \cite{ullah_2024_quantum}.

\section{Discussion}
To assess real-world applicability, the proposed framework can be implemented in pilot studies across multiple hospitals, ensuring compliance with healthcare privacy regulations. We envision deploying the system in collaboration with hospitals that maintain decentralized MRI datasets, allowing validation of model performance in practical settings. Furthermore, we propose testing our framework within federated cloud-based infrastructures, enabling scalability for large-scale medical AI applications.

\subsection{Privacy and Security}
The use of QKD-based encryption ensures robust protection of model weights during transmission, effectively mitigating risks associated with eavesdropping and gradient inversion attacks. The theoretically secure nature of QKD offers a significant advantage over classical encryption techniques, which are susceptible to advanced cyber threats. By leveraging the principles of quantum mechanics, the framework enhances trust in collaborative healthcare AI systems.

\subsection{Impact on Model Performance}
The encrypted model achieved identical accuracy to the baseline model, demonstrating that the integration of encryption does not compromise the predictive capabilities of the framework. Furthermore, the slight reduction in loss suggests that the federated aggregation process benefits from distributed training insights, enhancing model convergence. These findings affirm that advanced security measures can be incorporated without adversely affecting model efficiency.

\subsection{Practical Implications}
The proposed framework is particularly relevant for healthcare applications in low- and middle-income countries (LMICs), where data privacy regulations and resource constraints often hinder the adoption of advanced AI solutions. By enabling secure and privacy-preserving collaborative training, the framework democratizes access to cutting-edge dementia diagnostic tools, promoting equitable healthcare outcomes globally.

\section{Conclusion and Future Work}

This paper introduces a novel federated learning (FL) framework enhanced with quantum key distribution (QKD)-based encryption for secure and privacy-preserving dementia classification. By addressing critical challenges related to privacy and security, the proposed approach demonstrates substantial potential for revolutionizing healthcare artificial intelligence (AI). It enables secure collaboration across distributed healthcare data sources, allowing for the creation of high-quality models without compromising patient confidentiality.

\subsection{Key Contributions}
The primary contributions of this study are as follows:
\begin{itemize}
    \item The integration of QKD-based encryption into federated learning, providing a robust mechanism to safeguard model weights against data interception and unauthorized access, thereby enhancing privacy and security.
    \item The design of a federated learning framework specifically tailored for dementia classification, which ensures patient privacy while simultaneously maintaining high model performance, even in the presence of sensitive medical data.
    \item The experimental validation of the proposed framework, demonstrating its ability to balance enhanced security with efficient and effective model training across distributed healthcare institutions.
\end{itemize}

\subsection{Future Work}
Although the results of this study are promising, several directions warrant further investigation to improve and expand the framework's applicability:
\begin{itemize}
    \item \textbf{Scalability and Generalization:} Future research will focus on evaluating the scalability of the proposed framework across diverse healthcare datasets and institutions. This will assess the framework's generalizability to other medical conditions and the robustness of its security mechanisms in larger, more complex environments.
    \item \textbf{Performance Optimization:} Future work will aim to optimize the computational efficiency of QKD-based encryption, ensuring that the framework can be seamlessly deployed in environments with limited resources, such as smaller healthcare facilities or mobile devices.
    \item \textbf{Real-World Deployment:} Conducting pilot implementations of the framework in real clinical settings will provide valuable insights into practical challenges, including integration with existing healthcare systems, and will highlight areas for further optimization.
\end{itemize}

In conclusion, this work represents a significant advancement in the development of secure and privacy-preserving AI for healthcare. By combining quantum-inspired encryption with federated learning, the proposed framework offers a promising solution to critical challenges in healthcare data sharing. Its potential to extend beyond dementia classification makes it a valuable approach for a wide range of medical applications, paving the way for broader adoption in healthcare AI.

\bibliographystyle{IEEEtran}
\bibliography{QFL}

\end{document}